\documentclass[review]{elsarticle}

\usepackage{lineno,hyperref}
\usepackage{natbib}
\usepackage{graphicx}
\usepackage[utf8]{inputenc}
\modulolinenumbers[5]

\journal{New Astronomy}

\biboptions{authoryear}

\begin{document}

\begin{frontmatter}

\title{Discovery and Characterization of Kepler-36b}

\author{Eric Agol\fnref{guggenheim}}
\address{Department of Astronomy, University of Washington, Seattle, WA 98195}
\fntext[guggenheim]{Guggenheim Fellow}
\ead{agol@uw.edu}

\author{Joshua A. Carter}
\address{Citadel LLC, 131 S Dearborn St, Chicago, IL 60603}

\begin{abstract}
    We describe the circumstances that led to the discovery of Kepler-36b, and the subsequent characterization of its host planetary system. The Kepler-36 system is remarkable for its physical properties: the close separation of the planets,  the contrasting densities of the planets despite their proximity, and the short chaotic timescale.  Its discovery and characterization was also remarkable for the novelty of the detection technique and for the precise characterization due to the large transit-timing variations caused by the close proximity of the planets, as well as the precise stellar parameters due to asteroseismology.  This was the first multi-planet system whose transit data was processed using a fully consistent photometric-dynamical model, using population Markov Chain Monte Carlo techniques to precisely constrain system parameters. Amongst those parameters, the stellar density was found to be consistent with a complementary, concurrent asteroseismic analysis.  In a first, the 3D orientation of the planets was constrained from the lack of transit-duration variations.  The system yielded insights into the composition and evolution of short-period planet systems.  The denser planet appears to have an Earth-like composition, with uncertainties comparable to the highest precision rocky exoplanet measurements, and the planet densities foreshadowed the rocky/gaseous boundary.  The formation of this system remains a mystery, but should yield insights into the migration and evolution of compact exoplanet systems.
\end{abstract}


\end{frontmatter}


\section{Introduction}

During his first sabbatical in 2011, Eric Agol spent part of the summer at the Harvard-Smithsonian Center for Astrophysics as the guest of Professor David (Dave) Latham.  Dave had the foresight to situate him across the hall from Hubble Fellow Josh Carter as he knew that both Josh and Eric had similar interests in transiting exoplanets.  Josh was working on various projects related to the \textit{Kepler} spacecraft, such as the newly discovered circumbinary planet Kepler-16b \citep{Doyle2011}.  Initially, Eric and Josh worked on the impact of finite cadence on the measurement of transit timing parameters \citep[which was never published, but has been studied in detail by][]{Price2014}.  

During subsequent discussions, Josh brought up the question of whether the \textit{Kepler} pipeline might be missing transiting planets, their dips blurred by transit-timing variations \citep[TTVs;][]{MiraldaEscude2002,Schneider2003,Schneider2004,Agol2005,Holman2005}.  This issue was on his mind owing to the recent discovery of Kepler-16b, a circumbinary planet which shows transit timing variations of the order of several days, making the transits of this planet highly non-periodic, something Eric had thought about in prior work \citep{Agol2005}.  Josh found that a full photometric-dynamical (``photodynamic") model of the Kepler-16 system  was required to account for the aperiodic timings and durations of the eclipses due to the large reflex motion of the binary stars about their center of mass.  

In retrospect, the issue with detection, then, is that it is computationally prohibitive to fit every possible photodynamical model for a transiting planet system.  The parameter space of possibilities becomes prohibitively large when unknown, non-transiting planets and/or planets with undetected shallow transits are included which perturb the positions and velocities of the transiting bodies in the system.  Consequently, the standard approach of assuming periodic orbits when searching for transiting planets may be biased against systems which are strongly perturbed.  For example, the ``Box-fitting Least Squares" (\texttt{BLS}) approach, which assumes uniformly-spaced transits \citep{Kovacs2002}, may have less sensitivity to systems in which each transit is too shallow to be easily found with a pipeline or by eye (as most of the circumbinary planets had been found). If the transit timing variations are comparable to the duration of transit, then the transit signal is smeared when folded with a periodic ephemeris in \texttt{BLS}, which reduces the signal-to-noise.  This problem had been discussed in the literature by \citet{2011MNRAS.417L..16G}, but without a clear solution for the general case, although \citet{Ofir2008} suggested a solution for the circumbinary case.  Eric shared the same concern as Josh, but also lacked a solution.  

Josh shared some papers he had found about pulse detection within the electrical engineering literature, but they did not look too promising.  He mentioned the terms ``quasi-periodic" and ``pulse," and so later that evening Eric Googled the phrase ``quasi-periodic pulse detection."  The first item that appeared in the list of search results was a paper by \citet{Kelmanov2004} titled ``A posteriori joint detection and discrimination of pulses in a quasiperiodic pulse train," which he thought sounded rather promising, but upon skimming the paper, did not grasp precisely how it worked.  Eric emailed this off to Josh with the subject line ``this may be relevant," and Josh responded with an excited email stating that this seemed like a promising solution.  

\begin{figure}
    \centering
    \includegraphics[width=\hsize]{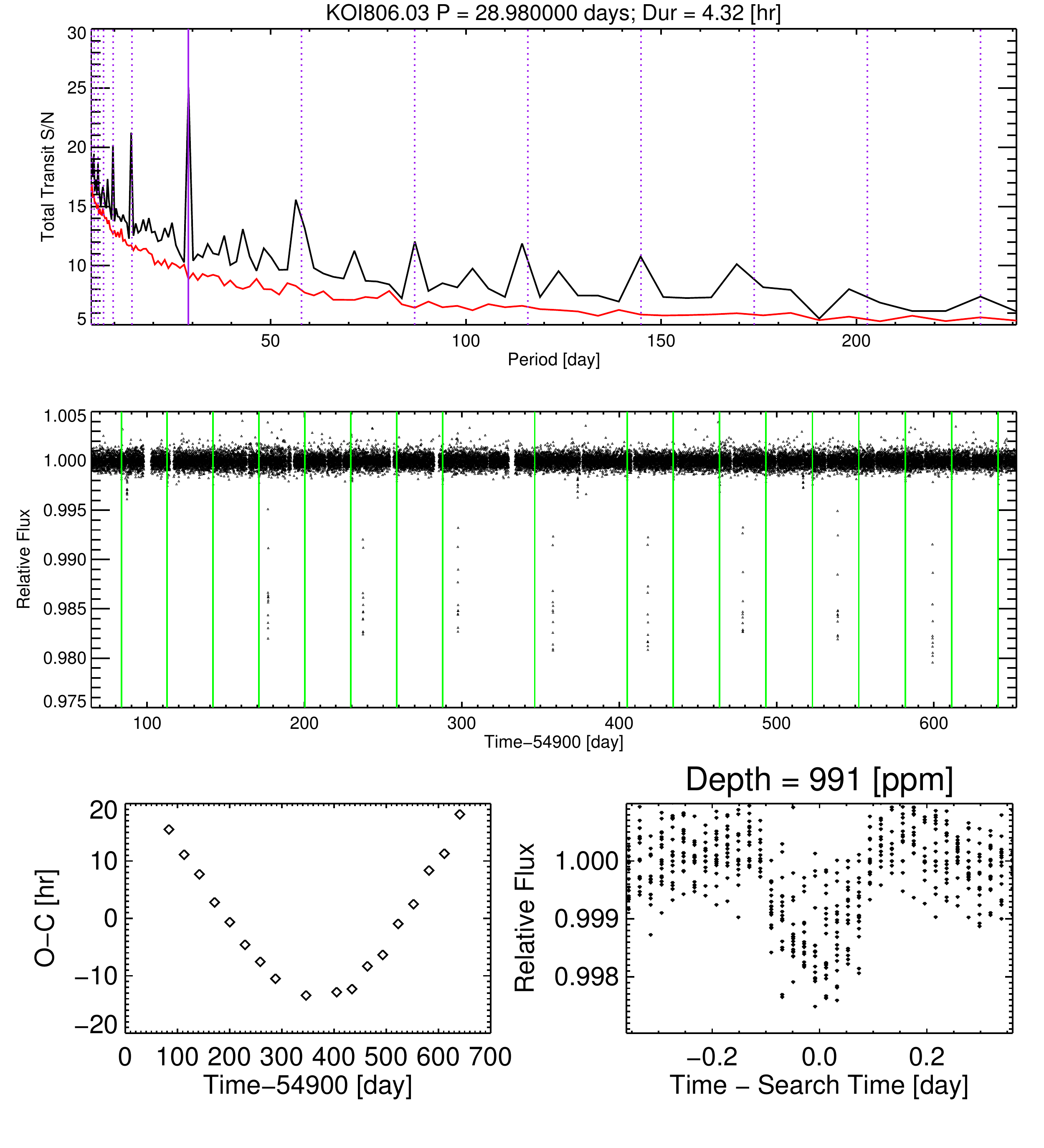}
    \caption{Detection of KOI-806.03 with the \texttt{QATS} algorithm.  Top panel: the \texttt{QATS} periodogram, which is the signal-to-noise of the \texttt{QATS} signal as a function of orbital period.  The peak period is indicated with a solid vertical line, while the dotted lines are harmonics.  Middle panel:  Light curve with the transits of planet 806.03 indicated.  Bottom left panel:  Transit times found by \texttt{QATS}, with a linear ephemeris subtracted (``O-C" = ``observed-calculated", or TTV).  Bottom right panel:  Folded transits, shifted to align the mid-point of transits.}
    \label{fig:KOI806.03}
\end{figure}

Josh quickly coded the technique up, and got it working within a day.  It took Eric a bit longer, but he also got it working with some help from Josh on the notation in the paper.  They tried it out on simulated and \textit{Kepler} transiting planet systems, and within a few days they were detecting quasi-periodic transiting exoplanets, and also used it to detect planetary transits by periodic \textit{Kepler} Objects of Interest as well.  One of the first planet systems they applied it to was KOI-806/Kepler-30 which has a shallow transit by the third planet candidate with extremely large transit-timing variations
(Figure \ref{fig:KOI806.03}), and the algorithm picked up on the planet with the correct period.  However, this was not the ideal test case as the planet candidate had already been identified by the \textit{Kepler} pipeline; with a depth of 1 mmag, it gave sufficiently high signal-to-noise to be detectable over timescales where the ephemeris could still be approximated as linear.   As a side note, this planet system has the distinction of being the first planet with a convincing detection of transit-timing variations from the ground \citep{Tingley2011}.

They discussed writing up the technique, and brainstormed a name for the algorithm.  Eric came up with the acroynym ``QATS" algorithm\footnote{Which stands for ``Quasi-periodic Automated Transit Search" algorithm}, and when he relayed this to Josh, Josh shared that he had just played the word ``QATS" in Scrabble the night before.  They settled on this acronym to describe the technique \citep{Carter2013}.

After Eric departed the CfA in August 2011, Josh continued to develop the \texttt{QATS} algorithm, and to develop a pipeline to apply it to the raw \textit{Kepler} dataset.  This was a significant amount of code development to carry out.  The \texttt{QATs} algorithm requires a uniform set of data, evenly spaced in time, with uniform, white noise, and zero average mean continuum (i.e.\ the flux outside of transit must be zero).  The \textit{Kepler} data have instrumental variations due to pointing fluctuations; gaps in the data every thirty days when the data were downlinked;  variability due to star spots, oscillations, and granulation; outliers due to cosmic rays; and other features such as electronic noise, asteroids, charge bleeding, etc.

Josh wrote a pipeline that carried out outlier rejection (typically cosmic rays cause positive deviations, which can be distinguished from transits), detrended the data using a running polynomial, and filled in missing data in data gaps to maintain the uniform cadence.  Fortunately the \textit{Kepler} team had the foresight to demand a uniform cadence throughout the lifetime of the mission, so the data gaps could be filled in with an integer number of cadences, which is a prerequisite for the simplest form of the \texttt{QATS} algorithm (otherwise interpolation would have been necessary).  All of this pre-processing then led to light curves which could be fed into the \texttt{QATS} algorithm, with one very minor last modification: \texttt{QATS} is a pulse detection algorithm, while the transits (with the background polynomial subtracted) are negative deviations in the flux of the star.  So the final step is to flip the transits in sign to become pulses.

Josh and Eric decided to focus their search on the host stars for which there were already detected planetary candidates, the so-called ``\textit{Kepler} Objects of Interest", or KOIs.  If one planet transits, the plane-parallel nature of planetary systems enhances the probability of another planet transiting \citep{Holman2005,Ragozzine2010}.  They discussed whether they should focus their search on planets for which there were already large apparent transit-timing variations.  However, they decided against this as it seemed intuitively unnecessary:  a larger planet would have a larger transit depth, and thus be more easily detected.  A smaller perturbing planet that might escape detection may have a smaller mass, and thus would perturb the larger planet less.  Thus, the larger planet might show smaller TTVs than the (undetected) smaller planet.  An example of this behavior is nicely demonstrated with Kepler-289, which has a giant planet, Kepler-289c, with small TTVs perturbed by a smaller mini-Neptune with large TTVs\footnote{Kepler-289d (aka ``Planet-Hunters 3") was initially missed by the \textit{Kepler} pipeline, and also missed by our initial \texttt{QATS} search due to the large amplitude star-spots of the host star;  the Planet Hunters crowd-sourced \textit{Kepler} data search found the additional planets in this system.} \citep{Schmitt2014b}.  

\begin{figure}
    \centering
    \includegraphics[width=\hsize]{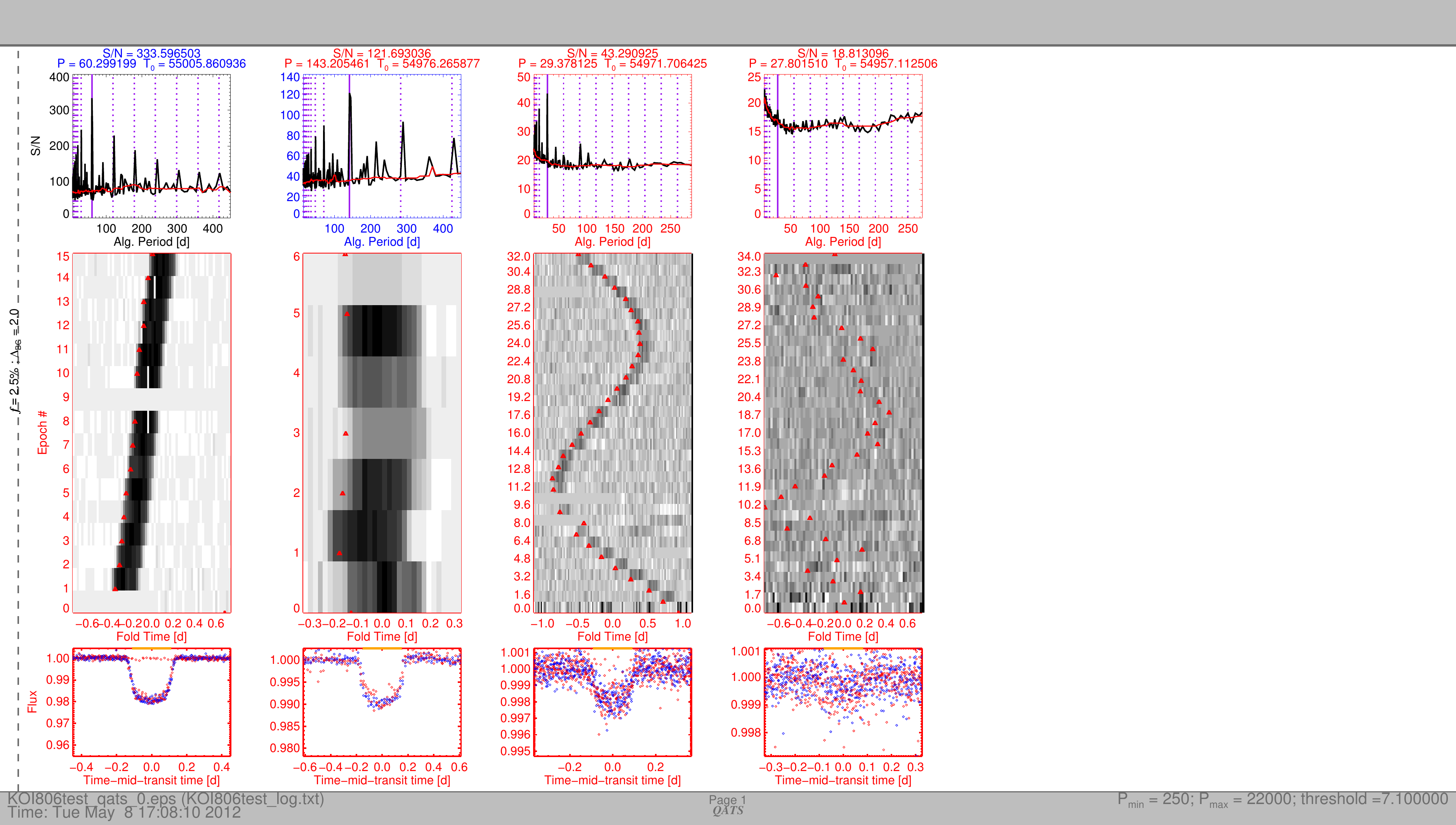}
    \caption{Repeated search, masking, and application of\texttt{QATS} to KOI-806.  This revealed the three planets in the system, shown in the first three grayscale image columns, with the third one being KOI-806.03 which has large TTVs.  The fourth column shows a fourth candidate that was rejected due to its proximity in period to the third planet and low significance.}
    \label{fig:KOI806_search}
\end{figure}

The search for each star proceeded by first letting \texttt{QATS} find the first known KOI, and then an extra step of pre-processing was required to mask the transits of the detected planet and fill in with cadences of zero signal.  Then, \texttt{QATS} was applied again to the masked light curve, searching for an additional planet.  This was followed by repeated masking and searching until no more significant signals were found.   An example of this process is shown in Figure \ref{fig:KOI806_search}.

\begin{figure}
    \centering
    \includegraphics[width=0.48\hsize]{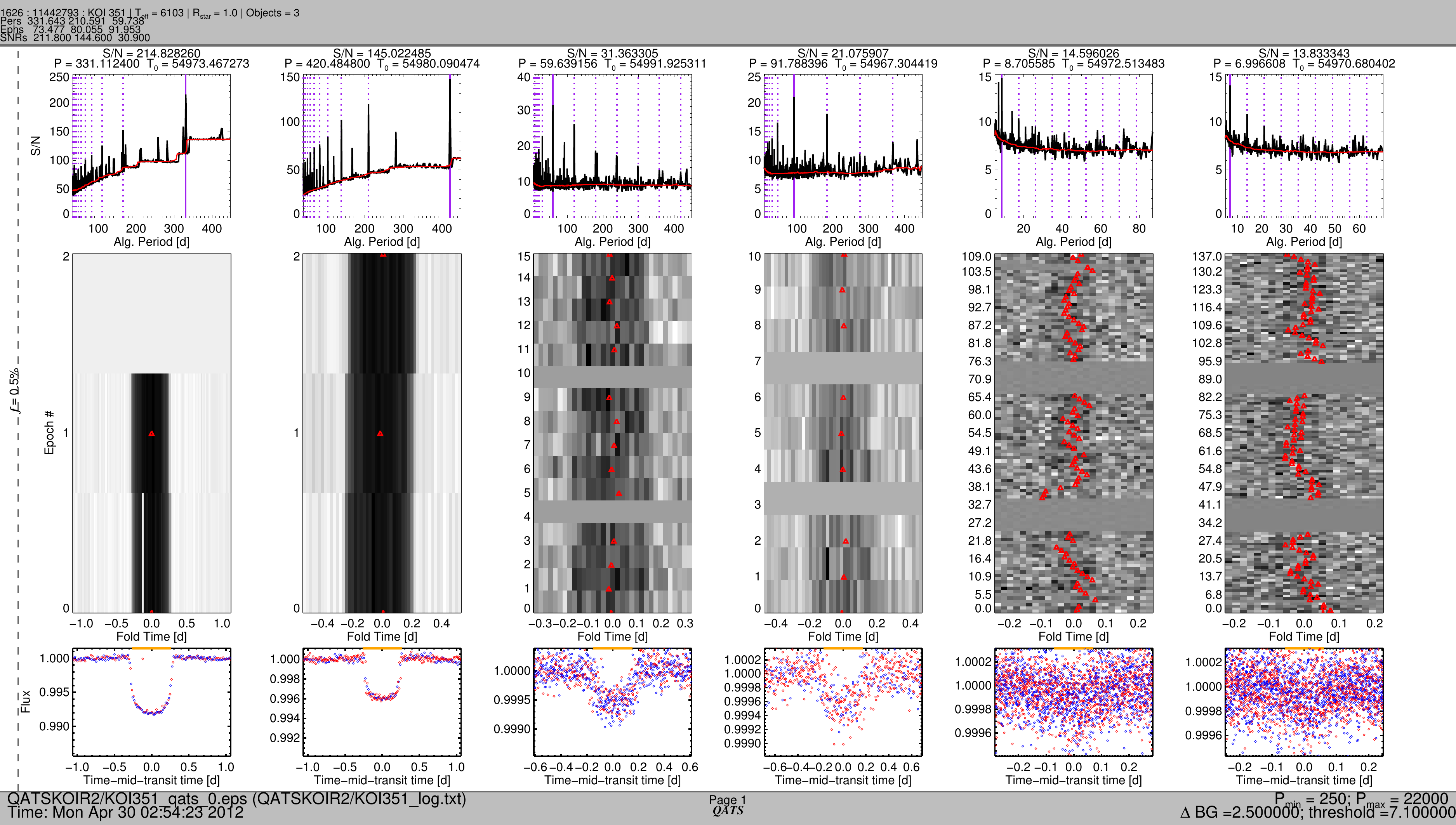}    
    \includegraphics[width=0.48\hsize]{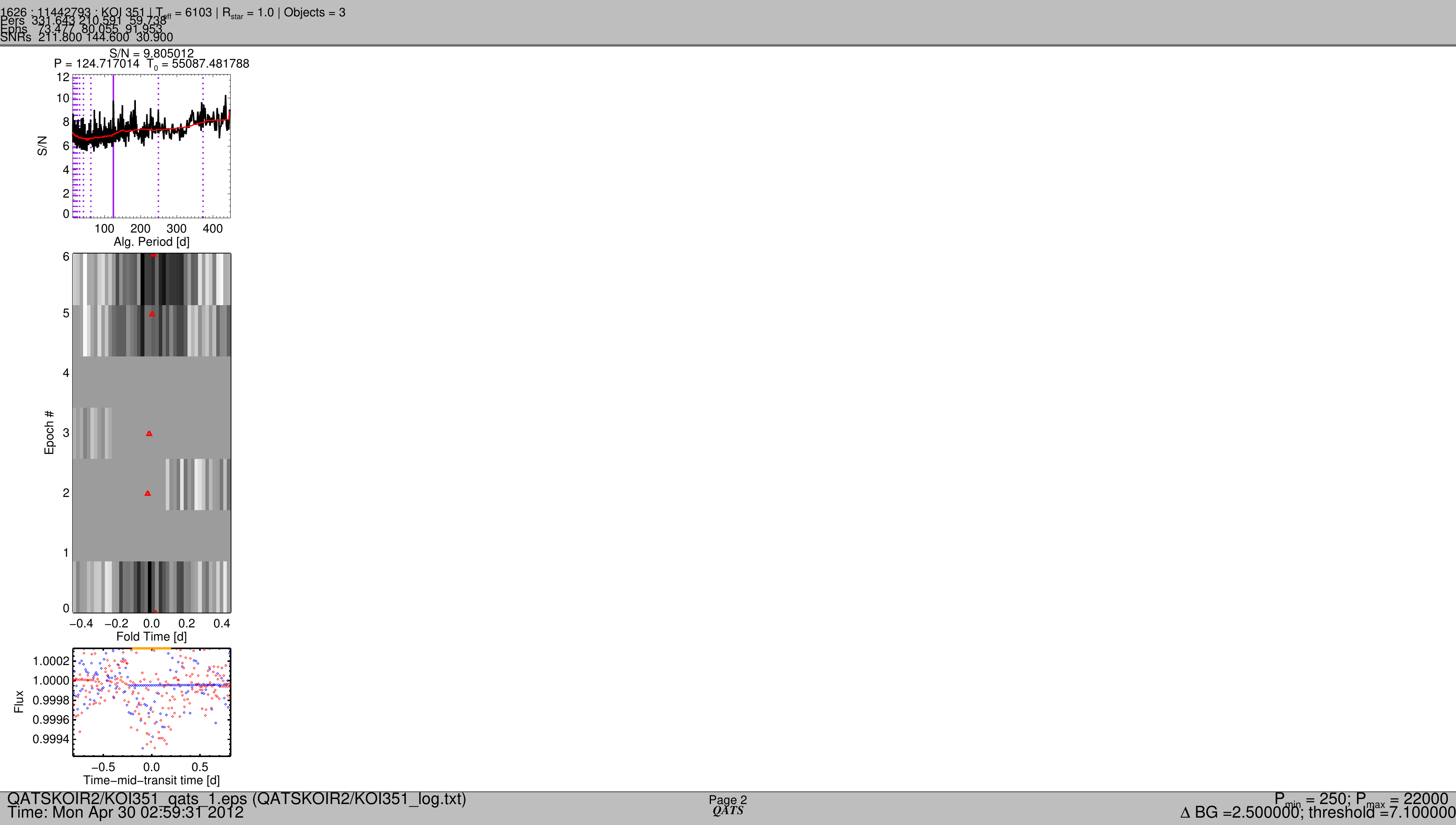}
\caption{Repeated search, masking, and application of\texttt{QATS} to KOI-351.  This revealed the seven planets in the system, several of which showed large TTVs, but none of which were large enough to warrant the \texttt{QATS} algorithm.}
    \label{fig:KOI351}
\end{figure}

This process ended up being very efficient in finding periodic KOIs;  in fact, the record holder was KOI-351, which showed seven planet candidates found with the \texttt{QATS} algorithm (Figure \ref{fig:KOI351}).  This system was announced in February 2013 at the Aspen conference ``Exoplanets in Multi-Body Systems in the \textit{Kepler} Era," at which point Aviv Ofir shared that he had found an eighth planet candidate in the system with a period of 14 days, albeit with a shorter transit duration than expected given this orbital period.  The first seven planets in this system ended up being discovered by the Planet Hunters team as well as a European team \citep{Schmitt2014a,Cabrera2013}, in addition to the \texttt{QATS} discovery which was reported along with other multi-planet systems by the \textit{Kepler} team \citep{Lissauer2014,Rowe2014}.  Later the eighth planet candidate was found again with machine learning \citep{Shallue2018}.  The discovery plots of KOI-351 are shown in Figure \ref{fig:KOI351};  several of the planets show TTVs, but none of which are large enough to have warranted the \texttt{QATS} algorithm.  

Josh completed the search of \textit{Kepler} Objects of Interest when one or more planets had been discovered to see if these had any additional planets that were quasi-periodic, and hence missed by the \textit{Kepler} pipeline planet search.  Eric didn't have access to the \textit{Kepler} data at that point as it was proprietary still, and he was not part of the \textit{Kepler} collaboration.  Josh wrapped up this search, found a couple of interesting candidates, but then turned back to working on projects he was behind on for the \textit{Kepler} collaboration (such as Kepler-16), while Eric turned back to other projects and teaching, awaiting approval to examine the proprietary \textit{Kepler} data.

In October 2011, Eric was granted collaborator status with the \textit{Kepler} team, thanks to the support of Dave Latham.  Josh shared the results of his KOI companion planet search with \texttt{QATS}, within which he had flagged ten objects of interest.  For KOI-277, he noted: 

\texttt{277 (This one has wild TTVs.  I kept it because Jason has in his notes
that 277.01 has big TTVs as well)}.\footnote{He was referring to either Jason Rowe or Jason Steffen.}

\begin{figure}
    \centering
   \includegraphics[width=\hsize]{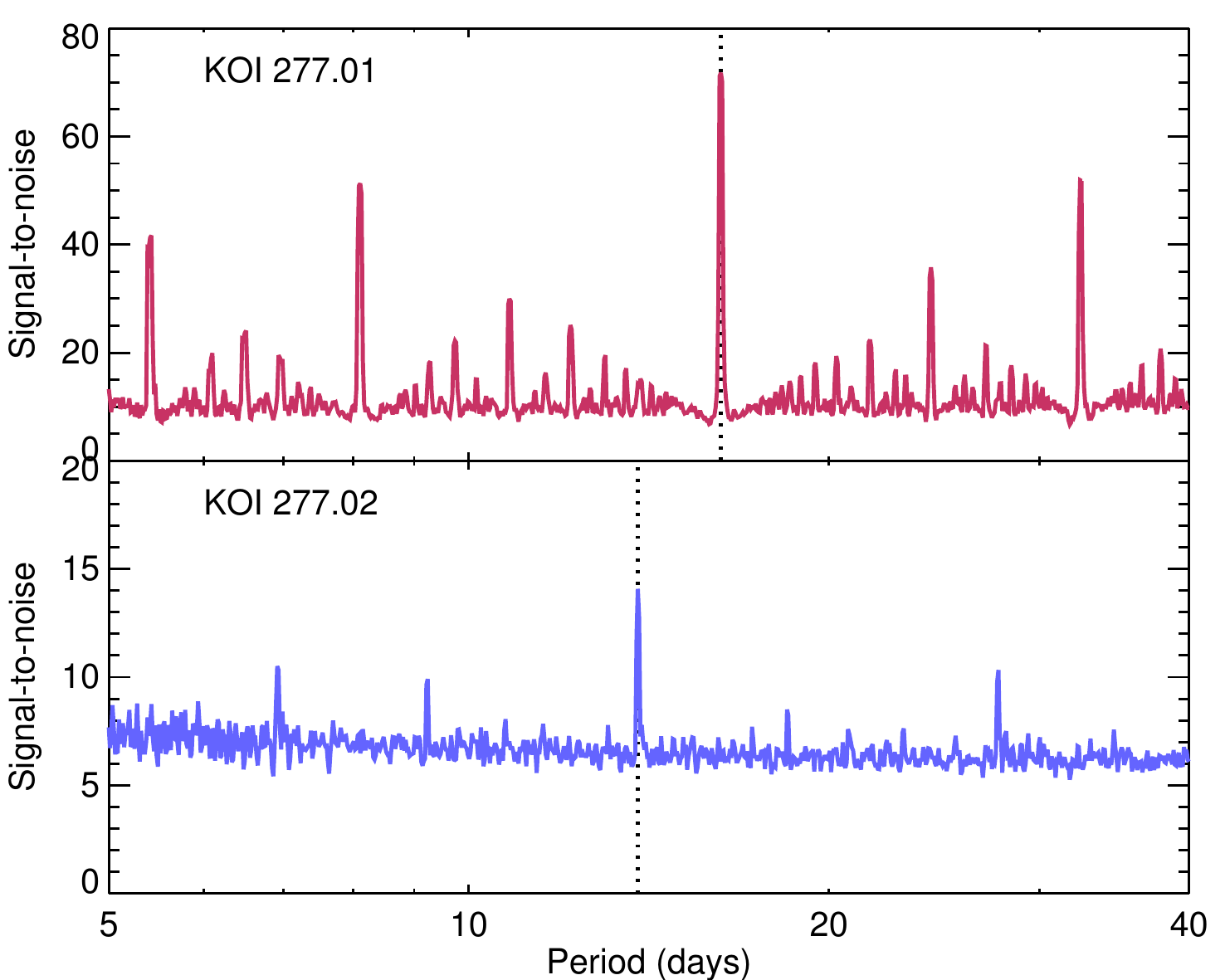}    

\caption{Repeated search, masking, and application of \texttt{QATS} to KOI-277.  A known planet, KOI-277.01 (36c), was recovered, while a new one, KOI-277.02 (36b), was revealed for the first time by the \texttt{QATS} algorithm.}
    \label{fig:KOI277}
\end{figure}

Kepler-277.01/Kepler-36c had been found earlier by \citet{Borucki2011}, which was the catalog that Josh drew upon to carry out the search for additional transiting companions. This candidate was found despite the large TTVs due to the large depth of transits;  it was flagged as a TTV candidate in one of the early catalogs \citep{Ford2011}.
Subsequently, at the \textit{Kepler} science conference in December 2011, Eric was scheduled to give a talk, in which he was planning to discuss the \texttt{QATS} algorithm.  He had just been granted collaborator status with the \textit{Kepler} team, and thought it would make a much better talk if \texttt{QATS} had detected an {\it actual} planet with the algorithm, rather than the simulated systems he had available to show. He perused Josh's search in detail, and of the new candidates that Josh had found, one of them stood out that showed large transit-timing variations:  KOI-277.02/36c (Fig.\ \ref{fig:KOI277}).  It is unclear whether \texttt{QATS} was required for the detection; later the planet was detected with the \texttt{BLS} algorithm \citep{Kovacs2002} by Aviv Ofir and Stefan Dreizler \citep{Ofir2013}.  The candidate was missing It was missing in the Q1-6 KOI paper with the first 16 months of Kepler data \citep{Batalha2013}, while it was included in the Q1-8 data paper with the first 22 months of Kepler data \citep{Burke2014}.  So, although \texttt{QATS} hastened the detection of the planet, it was not requisite for its discovery.

\begin{figure}
    \centering
    \includegraphics[width=\hsize]{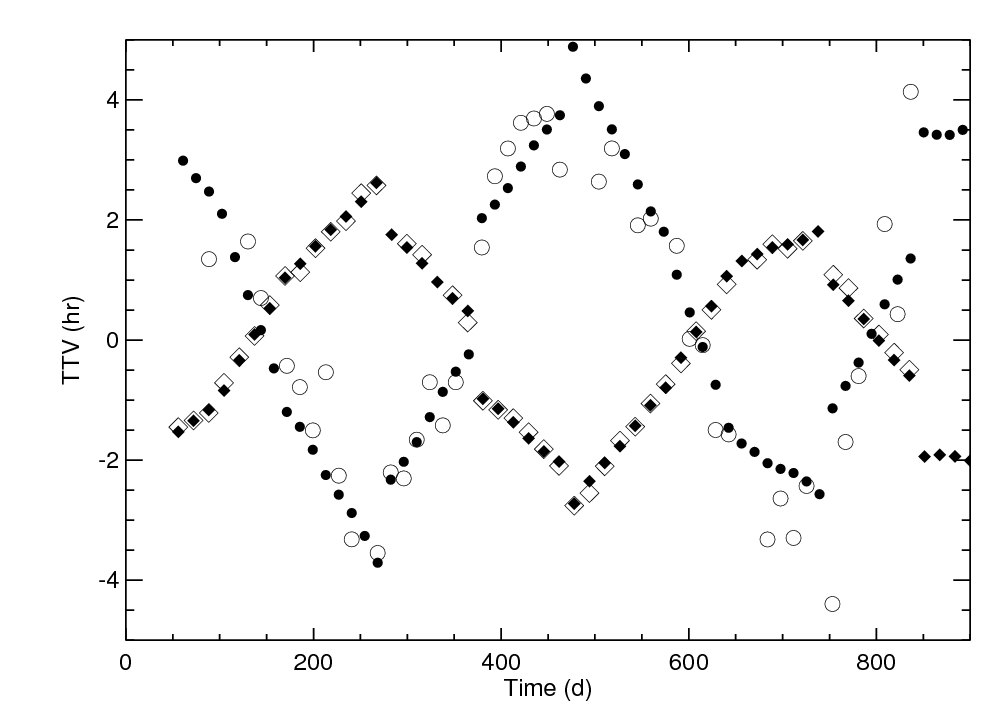}
    \caption{One of the initial fits to the transit-timing variations of KOI 277.02/36b (circles) and 277.01/36c (diamonds).  Open points are the data; solid points are the model.}
    \label{fig:initial_TTV}
\end{figure}

This system immediately caught Eric's attention, as well as the fact that the planets were very closely spaced in period, with a ratio close to 6:7.  In the last-ditched hopes of rescuing his conference talk, he spent a late night modeling the system with a dynamical integrator, and found an initial solution that looked promising (Fig.\ \ref{fig:initial_TTV}).  He excitedly shared this with the TTV team the next day at the conference, including Josh, Eric Ford, Jason Steffen, and Dan Fabrycky.  

With the detection of KOI-277.02/36b, the initial fits were challenging as the \texttt{QATS} algorithm did not yield precise times of transit for the planet candidate.  Each transit is so shallow that transit light curve fits had difficulty finding the individual time of transit.  Sometimes noise fluctuations would pull the fit off by times of order the transit duration.  However, it was apparent that the transits of 36c showed discontinuities in the ephemerides every six transits;  this coincided with conjunctions of the two planets based upon their ephemerides.  Josh carried out a fit to every seven transits of 36b in between the conjunctions - this gave sufficient signal-to-noise to yield piecewise ephemerides for the planet.  Eric found the first fits to these transit times;  one of the first fits is shown in Figure \ref{fig:initial_TTV}.  It was apparent from these initial fits that the TTVs of 36b were larger than that of 36c, which made sense since 36c created larger transit depths, and thus ought to be more massive.  Initial mass estimates gave $5.160 \pm 0.135 M_\oplus$ and  $9.33 \pm 0.205 M_\oplus$, which were fairly close to the final published values.  However, the reduced chi-square of the fits to 36b was large, so it was quickly realized that a different approach was needed. 

Since the signal-to-noise of each transit of Kepler-36b was rather poor (which is why it probably escaped detection by the \textit{Kepler} pipeline in the first place), we decided that we would have to write a photodynamical model to simultaneously constrain the dynamics and transit depths of both planets, much like Josh had done for the modelling of Kepler-16b.  Josh agreed that this was the best path forward, and so they both began the process of carrying out a parallel analysis of the system.  

One of the biggest surprises of this system was the close proximity of the planets, which begged the question of how these could have migrated through lower order resonances, as well as the large difference in transit depths despite similarly sized TTVs, which indicated that these planets had very different densities, one similar to Earth (277.02/36b), and one more similar to Neptune (277.01/36c).  Eric gave a talk on \texttt{QATS} at the \textit{Kepler} conference, and was not able to discuss KOI-277 in the talk (even though he hinted that an interesting result was to come during the question period), but the discovery of this system certainly made it much more exciting to be speaking on \texttt{QATS} given that it had successfully detected a new planet!  This helped make up for a lack of sleep due to spending much of the conference nights writing code to model the TTVs of this system.  

The two planets would eventually come to be known as Kepler-36b (KOI-277.02) and Kepler-36c (KOI-277.01), where the KOI number reflects the order in which they were discovered, KOI-277.01/36c with deep transits had been found by the \textit{Kepler} pipeline, while KOI-277.02/36b was first flagged as a planet candidate by Josh's \texttt{QATS} search \citep{Carter2012}.  The system was searched for additional planets, but none were found.

October through December were spent using every bit of free time coding up a photodynamical code.  Eric wrote his own photodynamical code, with the dynamics computed in \texttt{FORTRAN}, while the photometric model \citep{MandelAgol2002} was coded up in \texttt{IDL}.  Josh had already written a photodynamical model in \texttt{C++} for the Kepler-16 system, and we felt it would be worthwhile to have two codes to double-check our results, and for trying out different approaches to the analysis;  both of these aspects proved to be valuable.

As Eric was starting out on the analysis of this system, Josh finished up other projects, and modified a \texttt{C++} version of the code for analyzing the KOI-277/Kepler-36 system.  One of the challenges of this analysis was in carrying out the Markov-chain Monte Carlo (MCMC) analysis to determine the posterior distribution of system parameters.  The large number of free parameters which had to be fit and the expensive computational cost of evaluating the likelihood function made it challenging to run the Markov chain long enough to be able to get a well-converged posterior probability distribution for the system.  Josh introduced Eric to Differential-Evolution MCMC (which had been introduced to Josh by Eric Ford) and Josh was able to implement a parallel version that ran on 512 cores at once \citep{terBraak2008}.  Eric ran his models on a small number of cores, and so he had to seek more efficient means of parameterizing the system to speed up the markov chain analysis.

During the initial analysis of the system, they found that the eccentricities were strongly correlated.  They found a reparameterization that got rid of the correlation, and improved the convergence of the chains.  They discussed this with the Kepler-TTVs/multis science working group, and Dan Fabrycky pointed out that the correlation had to do with the orientation of the epicycles of the orbits of the two planets.  Since the orbital proximity of the two planets is close, and since the eccentricities are small, the epicyclic approximation is a good approximation for the system.  At the times of closest approach when the inner planet passes the outer planet, called conjunctions, the separation of the planets depends upon the amplitude and the orientation of their epicycles.  The planets perturb one another most strongly at conjunctions, since gravitational acceleration scales with inverse separation squared. Consequently, the shape of the TTVs depends most strongly on the proximity of the planets at conjunction.  The eccentricities also affect the time of transit, and since the planets each have a well-measured ephemeris, this causes a strong correlation between the eccentricities of the two planets that precisely matched the correlation that one expects from the epicyclic approximation.  The reparameterization removed this correlation, and improved the convergence of the markov chains.

\begin{figure}
    \centering
    \includegraphics[width=0.51\hsize]{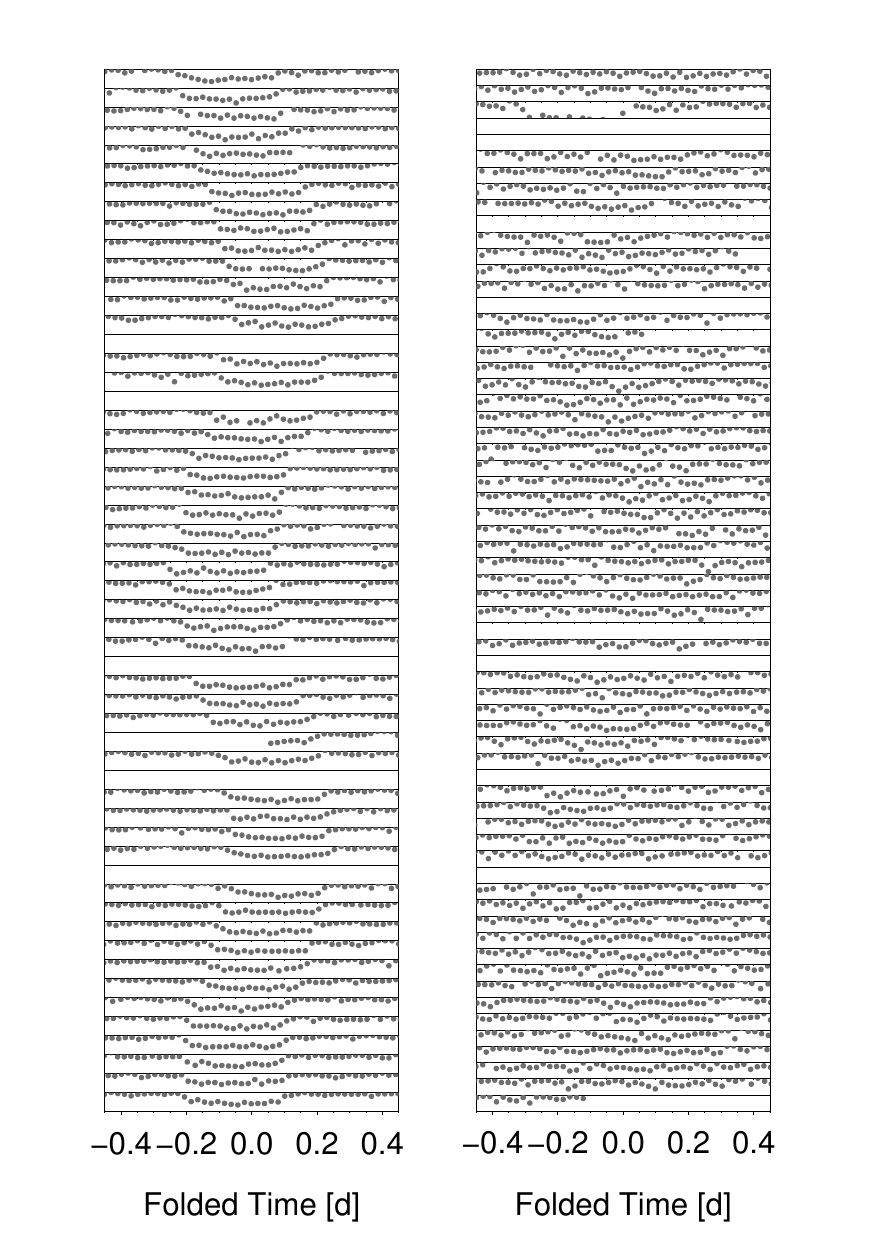}
    \includegraphics[width=0.46\hsize]{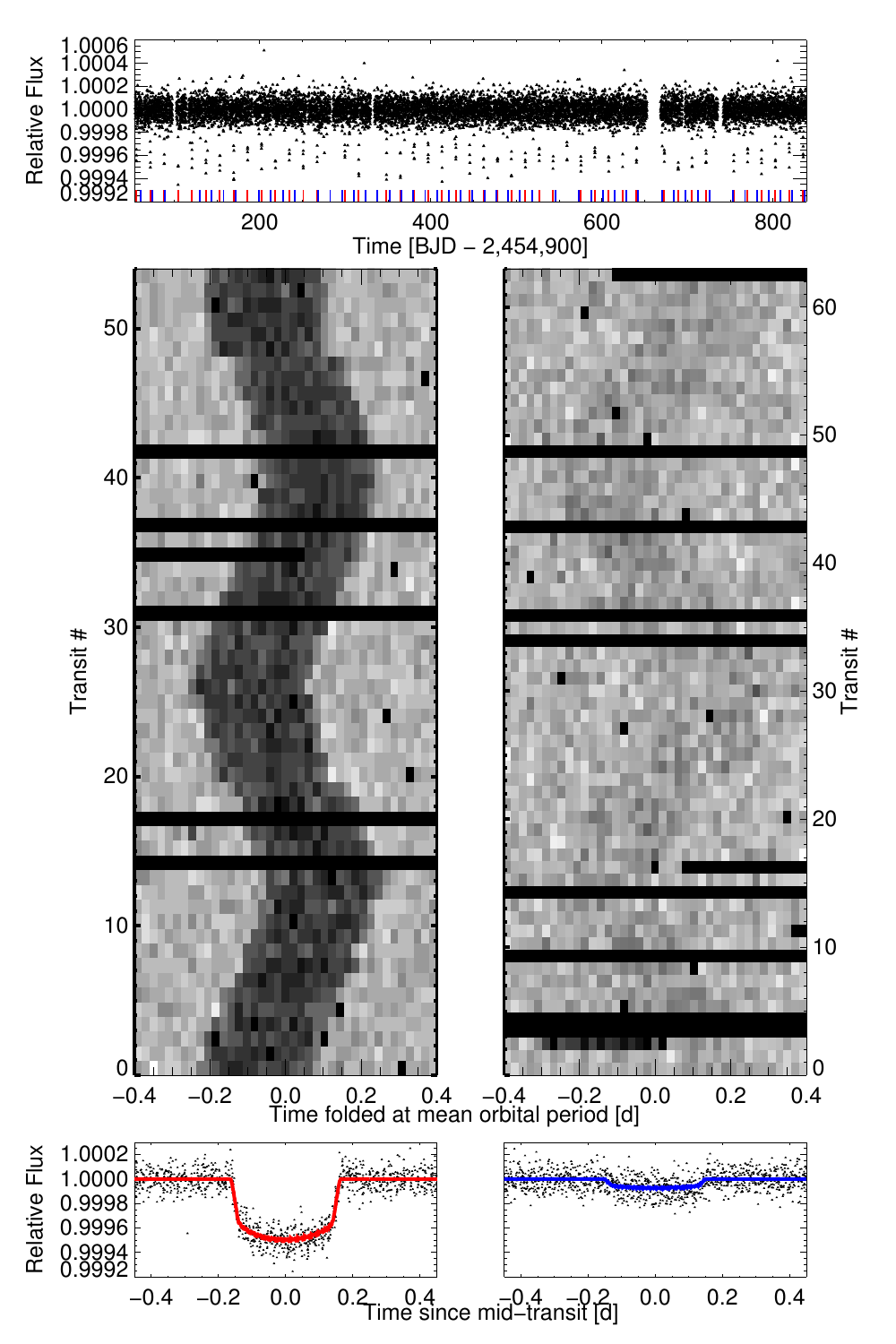}
    \caption{Some initial versions of the river plot, before the final green and blue version one was settled on for the paper.}
    \label{fig:riverplot}
\end{figure}

With the first photodynamical fits available, one issue was how to present the data in a clear fashion.  Dan Fabrycky suggested a river plot in which each transit was plotted, aligned to the expected time for the mean ephemeris, and each transit would appear shifted due to the transit-timing variations of the planets (Figure \ref{fig:riverplot}, left panel).  This approach worked well for Kepler-36c, but not for Kepler-36b as the smaller transit depth caused the noise to dominate, making the transits less apparent, and requiring a larger shift (relative to the transit depth) to avoid the noise from overlapping.  Eric realized that if each transit were plotted with a grey-scale, then they could be lined up in a grid, and the TTVs would cause a river plot that meandered back and forth as the transit times grew earlier and later (Figure \ref{fig:riverplot}, right panel).  Dan suggested that the grey scale be changed to a color scale with green at higher fluxes and blue at lower fluxes so that the riverplot would look like an actual river, resembling a SimCity\footnote{A city-building video game.} landscape.  This ended up being the final choice for the presentation of the data along with the photodynamical models in the paper, with missing data presented as a solid grey scale, and the ingress and egress times indicated with red tick marks.  Josh used the riverplot in later versions of his \texttt{QATS} KOI-search pipeline, which can be seen in some of the figures above.

\begin{figure}
    \centering
    \includegraphics[width=0.48\hsize]{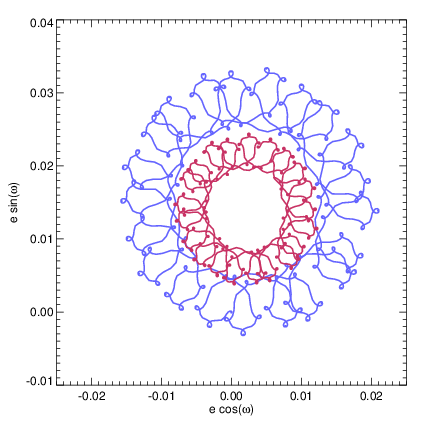}
    \includegraphics[width=0.48\hsize]{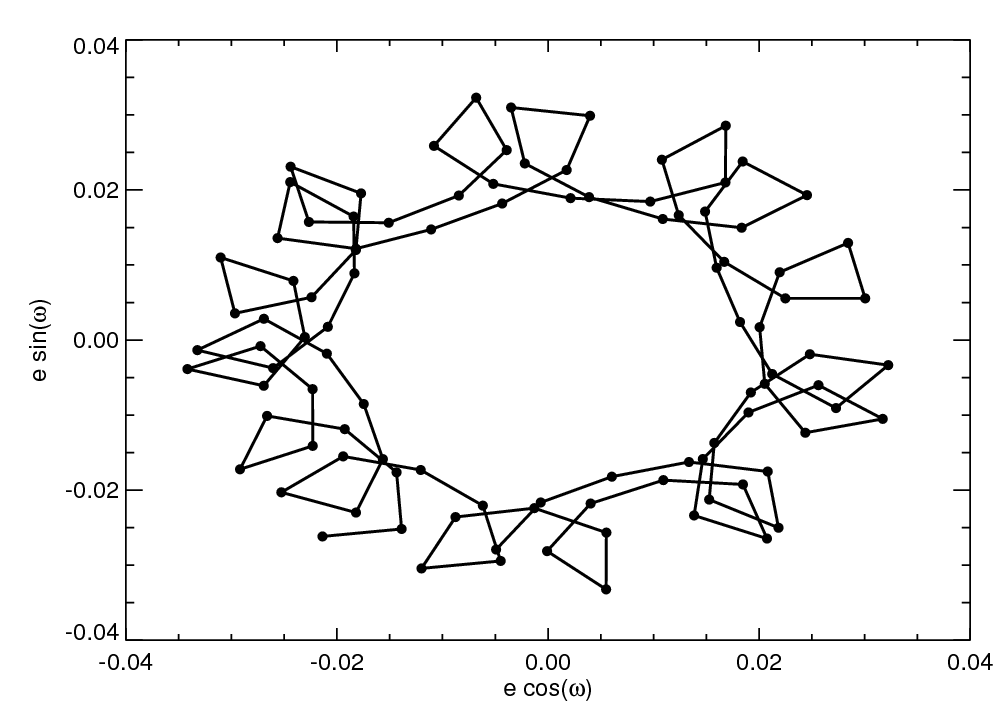}
    \caption{Evolution of the eccentricity vector for the two planets given the best-fit model (left); 36b is blue, while 36c is red.  Hill-approximation map of the evolution of the eccentricity vector of a model planet in the test-particle limit (right).}
    \label{fig:eccentric_evolution}
\end{figure}

Given the close proximity of the planets to one another, Dan Fabrycky suggested that the Hill approximation might make a good model for the planetary dynamics of the system.  He pointed out that a dynamical map might be a good approximation in which the planets followed Keplerian orbits in between conjunctions, and then received a kick at conjunction causing the orbital elements to change, as discussed in \citet{Duncan1989}.  He suggested that this matched the piecewise-continuous nature of the TTVs.  Indeed, it turns out that the eccentricity evolution of the system followed this model rather well.  Figure \ref{fig:eccentric_evolution} shows a comparison of the evolution of the eccentricity vectors of both planets for the best-fit model, which shows slow evolution between conjunctions, followed by fast evolution in the eccentricity, creating a petal shape which corresponds to the period of the TTV timescale.  Circulation of the eccentricity vectors on the secular timescale leads to a larger circulation of the eccentricity vectors, endowing a sunflower shape to the plots in which the circle, petals, and lines connecting the points represent the conjunction timescale, the super-period, and the secular timescale.

\begin{figure}
    \centering
    \includegraphics[width=\hsize]{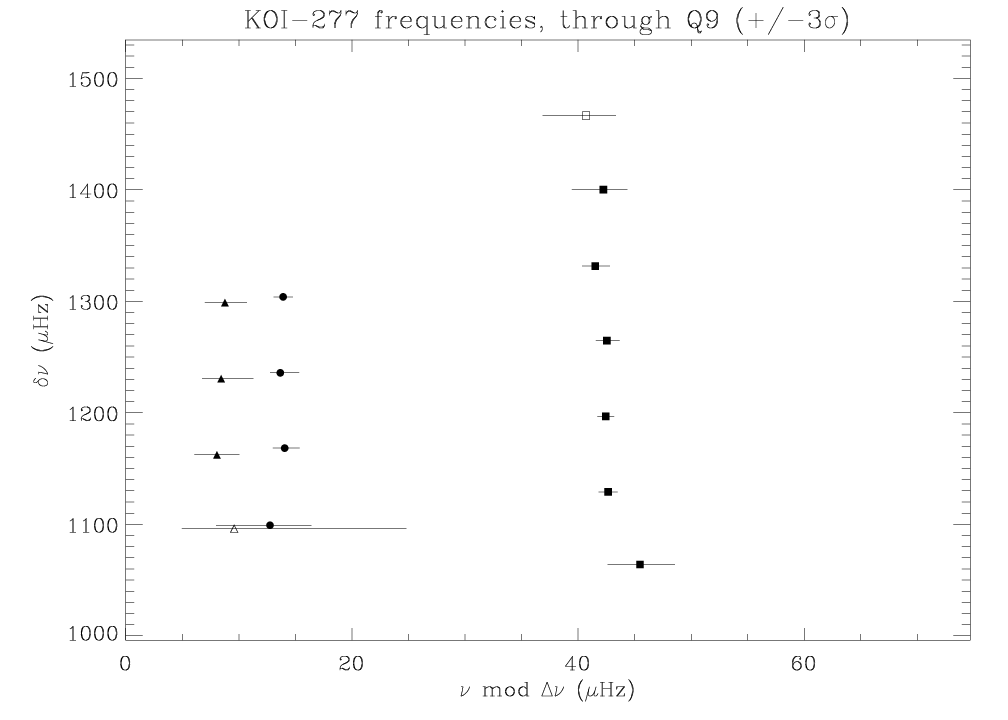}
    \caption{Asteroseismic analysis of Kepler-36. Credit: Bill Chaplin and KASC.}
    \label{fig:frequencies}
\end{figure}

With the photodynamical model maturing, it was realized that the precision of the planetary properties may well be limited by the knowledge of the stellar properties.  It was fortuitous, then, that the star was slightly evolved, and hence amenable to asteroseismic analysis of the short-cadence data which had been collected due to the TTVs identified in KOI-277.01/36c.  The \textit{Kepler} Asteroseismic Consortium (KASC), led by Bill Chaplin, was enlisted to carry out an asteroseismic analysis alongside the spectroscopic analysis to obtain precise system parameters.   A range of approaches were used to carry out the analysis of the oscillation frequencies to check for systematic errors and for consistency between the methods.  The star was sufficiently faint that the power spectrum was difficult to measure; nevertheless, the KASC team was able to measure the frequency of numerous oscillations, with which a precise characterization of the stellar mass, radius, and age was possible (Figure \ref{fig:frequencies}).  Since asteroseismology measures the oscillation frequencies with high accuracy, the technique is good at constraining the densities of stars since the density determines the dynamical timescale, $t_{dyn} \propto (G\rho)^{-1/2}$, which in turn determines the characteristic frequency spacing.  For Kepler-36, the asteroseismic density measurement yielded $\rho_{*,astero} = 0.3508 \pm 0.0056$ g/cc.  Likewise, the photodynamical model gave a constraint upon the density of the star as the durations of the transits, impact parameters, and orbital period can be combined with Kepler's law to yield a stellar-model-independent density of the star \citep{Seager2003};  there is a slight dependence upon the eccentricities of the planets which was constrained with the transit times and durations.  The photodynamic model yielded a completely independent measure of the stellar density of $\rho_{*,photo} = 0.3531 \pm 0.0053$, which agree at the 0.3$\sigma$ level.  This is remarkable consistency, and validates both techniques:  stellar oscillations and planetary transits can yield accurate and precise measures of stellar density.

Additional constraints upon the planetary properties were to come from the orbital stability analysis.  Katherine (Kat) Deck and Matthew (Matt) Holman showed that some of the posterior distribution would lead to orbits that were unstable in the long term;  this did not affect the planetary properties qualitatively, but did change the quantitative values.

With the measurement of precise stellar properties, the planet properties could be inferred with high precision.  In fact, the density of the star was much better constrained that either the mass or the radius, and since transit-timing variations measure the density-ratio of the planet to the star \citep{Agol2017}, then the planet densities were also better constrained than either their masses or radii.  This led us to plot the full posterior probability distribution of the planets in the mass-radius diagram, and since the iso-composition contours follow (approximately) along iso-density contours for 36b, this correlation was required for constraining the interior make-up of this planets.  The posterior distributions from the Markov chains were used directly by Leslie Rogers to place constraints upon the iron mass fraction of 36b ($M_{Fe}/M_{36b}=0.3\pm 0.1$, similar to Earth) and the volatile contents of both planets.  The surprise for these two planets was that their densities were so different despite having orbital distances that were so similar (the most similar of any two adjacent planets).  This led Eric Lopez and Jonathan Fortney to examine the mass loss history of the two planets, for which they found that the mass loss rate could be extremely sensitive to the core mass \citep{Lopez2013,Owen2016}.  Hence, the slightly larger mass core of 36c could retain a much more substantial volatile envelope, while 36c could be completely stripped.  Although this scenario contains parameterized physics, so far it has withstood scrutiny, and seems to be a plausible origin for the contrasting density of two planets in such close proximity.  The contrast in density of these planets presaged the transition in density between small planets and large \citep{Weiss2014,Rogers2015,Fulton2017}, with the transition between rocky planets and planets with gaseous envelopes occurring between 1.5-1.8 $R_\oplus$ attributed to atmospheric erosion \citep{Owen2017,Lehmer2017}.

The other puzzle for this system is how two planets could be formed, or evolve into, such close orbital proximity \citep{Rein2012}.  Mean-motion resonances might capture planets during migration, and prevent them from reaching such a close proximity, although disk turbulence might cause the planets to pass through these resonances \citep{Paardekooper2013}.  Another interesting scenario is that impacts by embryos may have caused the planets to migrate into their current configuration, and may have also stripped one of the planets via impact erosion, and possibly even caused the planets to swap positions \citep{Quillen2013}.
A more recent solution for the formation problem was found by \citet{Raymond2018}.  They find that migration of embryos, followed by merging, gave two planets with very different compositions - one rocky, and one icy - on very short-period orbits, in a similar configuration as Kepler-36b,c.  The inner planet originated from embryos within the snow line, while the outer planet originated from embryos beyond the snow line.  The two planets had outer companions which experienced outward migration, which may explain the lack of perturbing companions to these two planets.

Another constraint upon planet formation is the coplanarity of the planetary system.  The planets of the Solar System are extremely coplanar, with the exception of Mercury, which motivated the nebular hypothesis.  Strong dynamical interactions can lead to misaligned planets \citep{Juri2008}, and while the Rossiter-McLaughlin effect allows for the measurement of the misalignment of planet orbits with the host star rotation axis \citep{Gaudi2007}, the misalignment of planets is more difficult to gauge.  Planetary misalignment can cause the orbits to precess;  precession changes the transit path in front of the star, which can cause the duration of transits to vary.  The Kepler-36 planets did not show evidence of duration variations;  the photodynamical model could then be used to measure the mutual inclination of the planets' orbital angular momentum axes are aligned within $3.6^\circ$ at 3$-\sigma$ confidence.

Finally, the close proximity of the two planets reminded Matt Holman of the chaotic behavior of the closely spaced satellites of Saturn, Prometheus and Pandora \citep{Goldreich2003}.  He noted that the Lyapunov timescale could be short, which belied the long-term stability of the system;  however, bounded chaos could allow for stability despite the chaotic behavior.   This began a deeper dynamical exploration of the system, again using the posterior of the photodynamical fit, which revealed that the Lyapunov time was of order a decade \citep{Deck2012}.  Much of the posterior was unstable with longer-term integrations, which placed further constraints on the planetary parameters, and the role of high-order resonances, such as 34:29, was shown to play a role in the chaotic dynamics of the system (analagous to the 121:118 resonance affecting Prometheus and Pandora).

\begin{figure}
    \centering
    \includegraphics[width=\hsize]{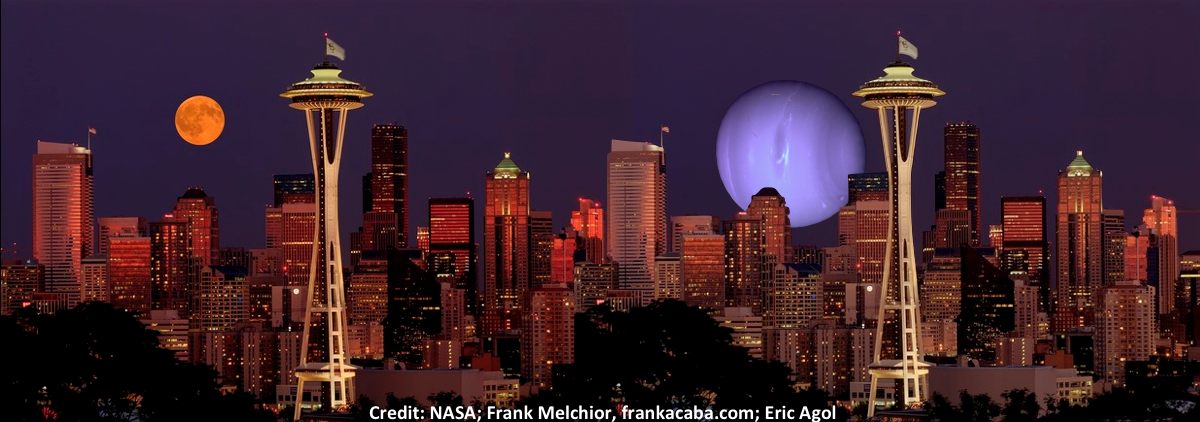}
    \includegraphics[width=\hsize]{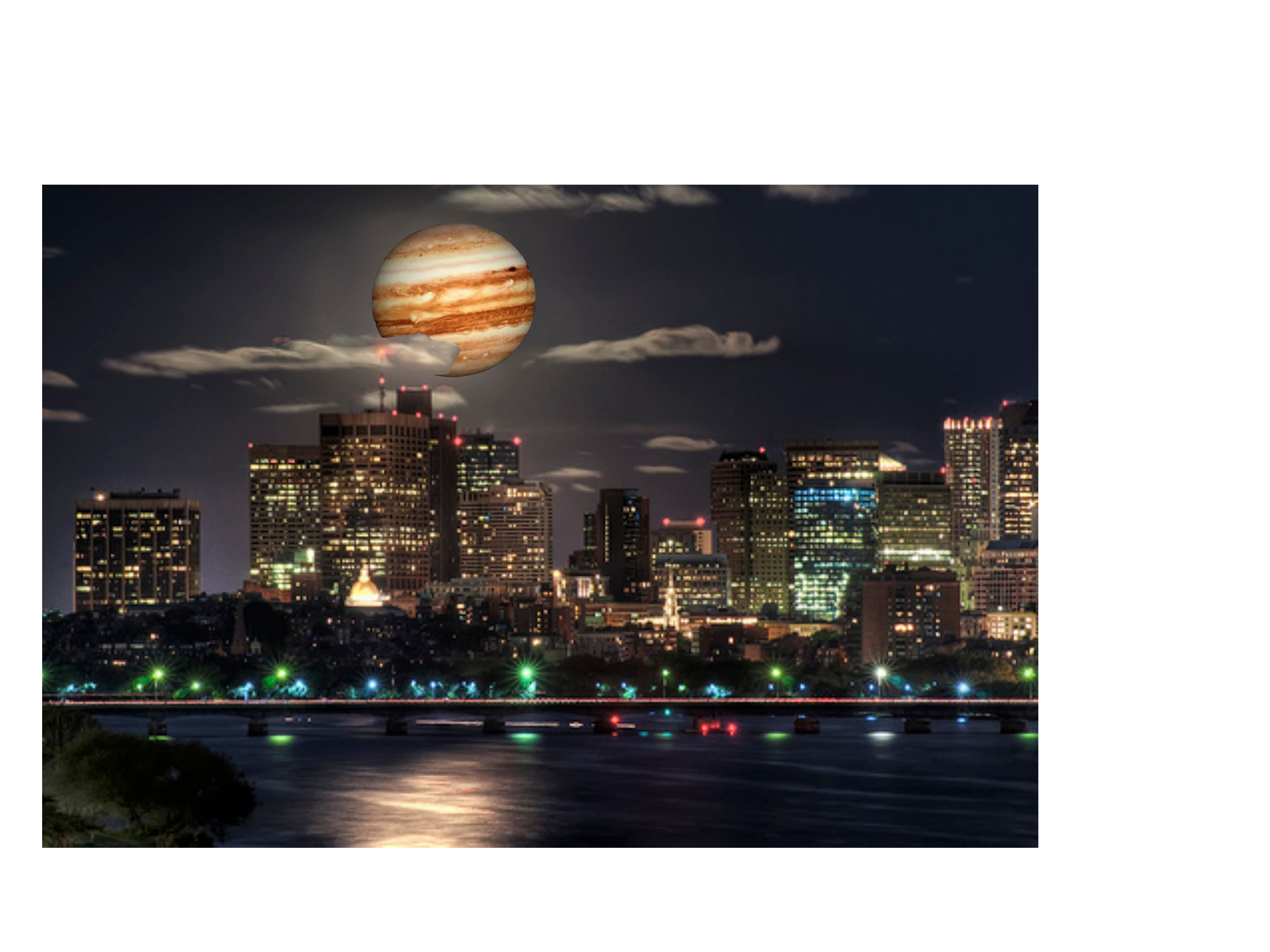}
    \caption{Visualization of Kepler-36c as seen from the surface of Kepler-36b at conjunction.}
    \label{fig:moon_Neptune_Seattle}
\end{figure}

After deriving the varied science results for this system, we had to consider how to present the results in a coherent fashion, highlighting the novelty of the system.  Given that multiple planet systems had been found with both super-Earths and mini-Neptunes, the novelty of this system was the close proximity in tandem with the contrasting densities, similar to the contrast between the greatest and least dense planets in our Solar system, Earth and Saturn, which have very different sizes and orbital distances.  The paper draft was started by Eric, and then handed off to Josh.  The final word-smithing of the paper was carried out by Josh Winn, who was able to clearly and concisely describe the scientific importance of the system and its novelty.  The other aspect of describing this system was how to present it to the general public.  The proximity of the planets would make the outer planet loom in the sky of an observer on the inner planet, spanning about three times the angular size of the Moon as seen from Earth.  To help visualize this, Eric found a picture of the moon over Seattle, and photoshopped it to replace the Moon with Neptune, which is similar in size to Kepler-36c; not to be outdone, Josh made a similar figure for the Boston skyline (Figure \ref{fig:moon_Neptune_Seattle}).  The result was used in the press release, and helped to convey how different the skies would be if one could visit such a planet.\footnote{Not to be outdone, Josh made a version showing the planet over Boston (a little bit of coastal rivalry).}

The planet Kepler-36b had the most precisely measured mass of a super-Earth at the time of its publication ($\approx$ 6\%).  Since that time two transiting rocky planets orbiting a much nearer K dwarf (at 6.5 pc) have had their masses measured with radial velocity to a higher fractional precision of 4-5\% \citep[HD 219134 b,c;][]{Gillon2017}.  However, given that the host star of Kepler-36b is slightly evolved, its characterization is amenable to the precision afforded by asteroseismology.  Thus, despite the more precise masses of HD 219134b,c, the {\it density} of Kepler-36b is known to $\approx 9$\%, which is better precision than any other known low-mass planet (HD 219134b,c have densities known to $\approx 11$\% precision).  This is somewhat ironic given that a larger star means a smaller transit depth for the same size planet, which is part of the reason Kepler-36b may have been missed initially.  Hence, the disadvantage of the size of the star for discovery of the planet turned out to be an advantage for characterizing the star, and hence more precisely characterizing the density and interior composition of the planet Kepler-36b, a feat which has yet to be matched.

\section{Acknowledgements}

E.A. acknowledges NSF grant AST-1615315 and 
the NASA Astrobiology Institute's Virtual Planetary Laboratory Lead Team,
funded through the NASA Astrobiology Institute under solicitation NNH12ZDA002C
and Cooperative Agreement Number NNA13AA93A.

\bibliography{kepler36}

\end{document}